\documentclass[aps,prl,superscriptaddress,twocolumn]{revtex4}
\usepackage{graphicx}
\newcommand{\jhu}
{Department of Physics and Astronomy, Johns Hopkins University, Baltimore, 
Maryland 21218, USA}
\newcommand{\rutgers}
{Department of Physics and Astronomy, Rutgers University, Piscataway, 
New Jersey 08854, USA}
\begin{document}
\title{Broken parity and a chiral ground state in the frustrated 
magnet CdCr$_2$O$_4$}
\author{Gia-Wei Chern}
\affiliation{\jhu} 
\author{C. J. Fennie}
\affiliation{\rutgers}
\author{O. Tchernyshyov}
\affiliation{\jhu}

\begin{abstract}
We present a model describing the lattice distortion and
incommensurate magnetic order in spinel CdCr$_2$O$_4$, a good
realization of the Heisenberg ``pyrochlore'' antiferromagnet.  The
magnetic frustration is relieved through the spin-Peierls distortion
of the lattice involving a phonon doublet with odd parity.  The
distortion stabilizes a collinear magnetic order with the propagation
wavevector ${\bf q}=2\pi(0,0,1)$.  The lack of inversion symmetry makes
the crystal structure chiral.  The handedness is transferred to
magnetic order by the relativistic spin-orbit coupling: the collinear
state is twisted into a long spiral with the spins in the $ac$ plane
and ${\bf q}$ shifted to $2\pi(0,\delta,1)$.
\end{abstract}

\maketitle

Frustration, defined as the presence of competing interactions, often
leads to unusual effects in magnets, particularly when it is combined
with a high symmetry \cite{Moessner06}.  A case in point is the antiferromagnet on the
``pyrochlore'' lattice that has a very high degeneracy of the ground
state if magnetic interactions are restricted to nearest neighbors
\cite{Moessner96}.  At the lowest temperatures such a magnet is
expected to retain a finite entropy per unit volume, as was indeed
observed in a group of pyrochlore magnets known as ``spin ice''
\cite{spin-ice}.  It is also well known that frustrated magnets with a
large degeneracy are prone to lattice distortions that reduce the
frustration by lowering the symmetry \cite{Yamashita00,OT02a}.  This
effect was observed in antiferromagnetic spinel ZnCr$_2$O$_4$
\cite{Lee00,Sushkov05}.  Unfortunately, the distortion in this
compound is rather intricate \cite{Ueda05}: it involves at least four
phonon modes with wavenumbers $2\pi\{\frac{1}{2},\frac{1}{2},\frac{1}{2}\}$.
Magnetic order in the distorted lattice is even more
complex: the magnetic unit cell is said to contain as many as 64 spins
\cite{Broholm}.  As a result, the basic story of a flexible pyrochlore
\cite{OT02b} does not apply to ZnCr$_2$O$_4$ and thus remains to be
fully tested.  Recent experimental characterization of another
spinel CdCr$_2$O$_4$ by Chung {\em et al.} \cite{Chung05} presents us
with an opportunity to do so.

Spinels $A$Cr$_2$O$_4$ with various nonmagnetic ions on the $A$ sites
are nearly ideal $S=3/2$ Heisenberg antiferromagnets with 
nearest-neighbor exchange on the highly frustrated ``pyrochlore''
lattice \cite{Lee00,Chung05,Ueda06}.  The size of the nonmagnetic ion
determines the Cr--Cr distance and thereby the strength of
exchange: 4.5 meV for Zn \cite{Lee00}, 1 meV for Cd
\cite{Chung05}, and a fraction of a meV for Hg \cite{Ueda06}.

CdCr$_2$O$_4$ undergoes a spin-Peierls-like lattice distortion
\cite{Yamashita00,OT02a} at $T_c = 7.8$ K \cite{Chung05}.  The lattice
symmetry is lowered from cubic ($Fd\bar{3}m$) to tetragonal (exact
space group unknown) with lattice constants $a = b \neq c$.  The unit
cell is elongated: $(c-a)/c \approx 5 \times 10^{-3}$.  In contrast,
the lattice is flattened, $(c-a)/c \approx -1.5 \times 10^{-3}$, in
ZnCr$_2$O$_4$.  It is remarkable that distortions in two very
similar compounds have opposite signs.  It is also surprising that the
magnitude of the distortion is larger in the compound with weaker
magnetic interactions.  This happens, apparently, because the 
quantity $(c-a)/c$ measures the uniform part of the
distortion only.  There are indications \cite{Lee00,Sushkov05} that
the nonuniform distortions in ZnCr$_2$O$_4$ are much larger than
the uniform component.  We will work under the assumption that the
lattice distortion in CdCr$_2$O$_4$ lowers the point-group symmetry of
the lattice but leaves the translational symmetry intact \cite{OT02b}.

The spins in CdCr$_2$O$_4$ remain disordered well below the
Curie-Weiss temperature and order simultaneously with the distortion.
Chung {\em et al.} \cite{Chung05} interpret the magnetic order as an
incommensurate spiral with the wavevector $\mathbf q = 2\pi (0,
\delta, 1)$ and the magnetization in the $ac$ plane.  They
offer two ordered structures compatible with the magnetic Bragg peaks.
In one the spins on every tetrahedron are nearly orthogonal (say,
close to directions $+\hat\mathbf{x}$, $+\hat\mathbf{z}$,
$-\hat\mathbf{x}$, and $-\hat\mathbf{z}$ on some tetrahedra), in the
other they are nearly collinear (say, two along $+\hat\mathbf{x}$ and
two along $-\hat\mathbf{x}$).  Since $\delta \approx 0.09$ is small,
we may treat it as an effect of a weak perturbation and begin our 
analysis at the commensurate point $\delta = 0$.

The Landau theory of a deformable ``pyrochlore'' antiferromagnet
\cite{OT02b} yields a variety of magnetically ordered states, with the
proposed orthogonal and collinear states among them.  Thus, from the
symmetry viewpoint, both candidate orders are plausible.  However, a
more ``microscopic'' treatment based on the actual physics of the
spin-lattice coupling (and as we will see, first-principles total energy
calculations)  invariably yields collinear ground states.
Stabilization of orthogonal ground states requires fairly exotic
interactions, such as a four-spin exchange {\em strongly} coupled to
the lattice \cite{OT02b}.  We therefore abandon the orthogonal state
and work with the collinear one. 

To simplify the calculations, we assume a clear
separation of relevant energy scales.  We treat the nearest-neighbor
Heisenberg exchange as the strongest interaction; its minimization
requires that the total spin of every tetrahedron be zero, which still
leaves a high-dimensional continuum of ground states
\cite{Moessner96}.  A weaker spin-lattice coupling selects from this
continuum a collinear ground state.  The weakest Dzyaloshinskii-Moriya 
(DM) interaction induces a slight misalignment of the spins and---in 
the presence of parity breaking---generates a spiral with a long period.
In this paper we outline our findings postponing a detailed
account to a future publication \cite{Chern07}.

{\em Starting point: commensurate collinear state.}  In a flexible
pyrochlore antiferromagnet, the spin-lattice coupling is an efficient
way to relieve spin frustration \cite{Yamashita00,OT02a}.  The
magnetoelastic coupling arises from the dependence of exchange on the
ion displacements $x_{\alpha}$ of spins: $E_\mathrm{me} = (\partial
J_{ij}/\partial x_{\alpha})\,({\bf S}_i\cdot{\bf S}_j)\,x_{\alpha}$.
Integrating out the phonons generates an effective biquadratic
exchange $-\sum_{i,j}(\mathbf S_i \cdot \mathbf S_j)^2$ favoring
collinear ground states.  Alternatively, this interaction can be
written in terms of the magnetoelastic force ${\bf f}=(f_1,\,f_2)$
whose components are linear combinations of the bond variables ${\bf
S}_i\cdot{\bf S}_j$ transforming as an irreducible doublet $E$ of the
tetrahedral group $T_d$ \cite{OT02b}. For a single tetrahedron, the 
magnetoelastic energy is $-{J'}^2|\mathbf f|^2/2k$, where $J'$ is a 
derivative of the exchange with respect to ionic coordinates and $k$ 
is the elastic constant of the vibrational doublet $E$.  The energy is 
lowest in a state with a tetragonal distortion, two weak and four 
strong bonds, and collinear spins \cite{OT02a}.  

\begin{figure}
\includegraphics[width=0.95\columnwidth]{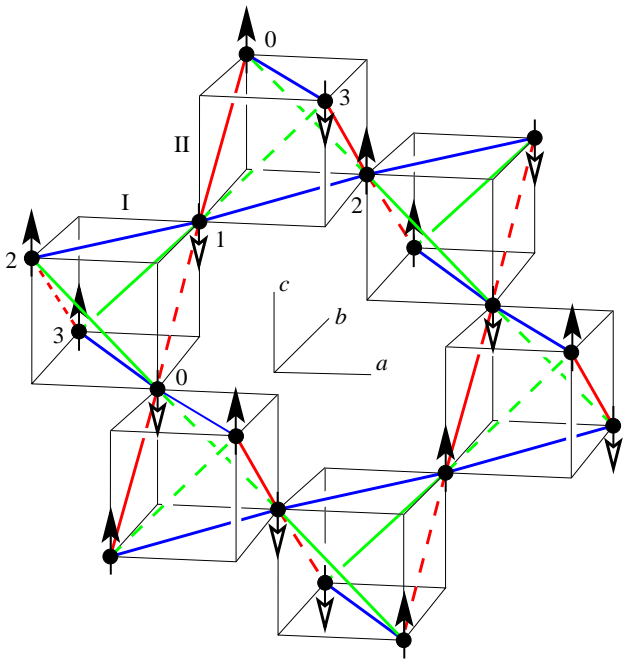}
\caption{\label{fig:redgreen} (Color online)
The collinear ground state stabilized by
the $\mathbf q=0$, $E_u$ phonon (adapted from Fig.~6 of
Ref. \onlinecite{OT02b}).  Tetrahedra of type I and II are flattened
along the $a$ and $b$ directions, respectively.  Solid (dashed) lines
satisfied (frustrated) bonds.  Frustrated bonds form spirals of the 
same handedness (left-handed in this case).  }
\end{figure}

When a distortion preserves the translational symmetry of the crystal,
generalization to an infinte lattice is straightforward \cite{OT02b}.
The existence of inequivalent tetrahedra with two different
orientations (I and II in Fig.~\ref{fig:redgreen}) adds inversion to
the symmetry group enlarging it to $T_d \otimes I = O_h$.  The
magnetoelastic energy of a primitive unit cell (four Cr ions) is 
\begin{equation}
  \label{eq:Eme}
  E_{me} = -K_g |\mathbf g|^2/4 - K_u |\mathbf u|^2/4,
\end{equation}
where $\mathbf g = \mathbf f^\mathrm{I} + \mathbf f^\mathrm{II}$ and
$\mathbf u = \mathbf f^\mathrm{I} - \mathbf f^\mathrm{II}$ are the
even and odd doublets of bond variables whose coupling constants are 
$K_{g,u}={J'}^2/k_{g,u}$, where $k_g$ and $k_u$ are the elastic constants 
of the even and odd
distortion doublets.  For $K_g > K_u$ the lattice undergoes a uniform
tetragonal distortion with $a=b>c$; the space group is $I4_1/amd$.
For $K_g < K_u$ the distortion has both even and odd components:
tetrahedra of types I and II are flattened along the $a$ and $b$
directions, respectively; the lattice is elongated overall, $a=b<c$;
the space group, $I4_{1}22$, lacks the inversion symmetry.
In both cases the ground states are collinear (Figs. 5 and 6 
in Ref. \onlinecite{OT02b}).

The antiferromagnetic order on the pyrochlore lattice can be described by
three staggered magnetizations ${\bf L}_i$ defined on tetrahedra of
type I: ${\bf L}_1 = ({\bf S}_0+{\bf S}_1-{\bf S}_2-{\bf S}_3)/4S$ and
so on.  The vanishing of the total spin of a tetrahedron in a ground
state, ${\bf M}^\mathrm{I}=\sum_{i=0}^{3}{\bf S}_i=0$, makes the three
N\'eel vectors ${\bf L}_i$ orthogonal to each other and imposes a
constraint on their lengths: $\sum_{i=1}^{3} L_i^2=1$.  In the state
shown in Fig.~\ref{fig:redgreen}, ${\bf L}_2={\bf L}_3=0$
and ${\bf L}_1=\hat{\bf n}_1\, e^{i{\bf q}\cdot{\bf r}}$ with 
${\bf q}=2\pi (0, 0, 1)$; $\hat{\bf n}_1$ is an arbitrary
unit vector.  This is also consistent with the data \cite{Chung05} if
we take the commensurate limit $\delta \to 0$.

{\em Long-period spiral.}  Spiral magnetic order can arise when
competing interactions destabilize a collinear ground state.  That can
happen when, e.g., the second-neighbor exchange is comparable to the
nearest-neighbor one.  However, further-neighbor exchanges are rather
weak in spinels $A$Cr$_2$O$_4$ and we have checked that they do
not destabilize collinear order (see below).

Alternatively, spiral magnetic order may reflect a chiral nature of
the underlying lattice.  The handedness is transferred from the
lattice to the spins by the relativistic spin-orbit coupling $\alpha
(\mathbf L \cdot \mathbf S)$.  Cubic spinels are non-chiral: the space
group $Fd\bar{3}m$ includes inversion.  However, parity is broken 
in the presence of the odd distortion $E_u$.  A chiral nature of the
distorted lattice becomes evident if one examines the locations of
frustrated bonds shown as dashed lines in Fig.~\ref{fig:redgreen}:
they form spirals of the same handedness.  Since the symmetry breaking
is spontaneous, experiments should reveal both right and left-handed
magnetic spirals originating in different domains.

In a Heisenberg magnet the spin-orbit interaction is manifested as 
the DM term ${\bf D}_{ij}\,\cdot\, [{\bf S}_i \times{\bf S}_j]$ 
\cite{Moriya}.  Elhajal {\em et al.} 
\cite{Elhajal05} have determined the vectors ${\bf D}_{ij}$ for the
``pyrochlore'' lattice up to a multiplicative constant.  In a single
tetrahedron, the DM term is
\begin{equation}
  E_\mathrm{DM}=-DS^2\, (\hat{\bf a}\cdot\,{\bf L}_2\times{\bf L}_3
  +\hat{\bf b}\cdot\,{\bf L}_3\times{\bf L}_1
  +\hat{\bf c}\cdot\,{\bf L}_1\times{\bf L}_2).
\end{equation}

Using the commensurate state as a starting point we parametrize the
magnetic structure as 
\begin{equation}
\mathbf L_i (\mathbf r) = e^{i \mathbf{q \cdot r}} \phi_i(\mathbf r)
\hat{\mathbf n}_i(\mathbf r),
\label{eq:LA}
\end{equation}
where $\hat{\mathbf n}_i(\mathbf r)$ and $\phi_i(\mathbf r)$ are the
directions and magnitudes of the staggered magnetizations.  (Note that
the three unit vectors $\hat{\mathbf n}_i(\mathbf r)$ are mutually
orthogonal.)  These parameters vary slowly in space.  Proximity to the
collinear state means that $\phi_2$ and $\phi_3$ are
small, while $\phi_1 \approx 1 - (\phi_2^2 + \phi_3^2)/2$.  The
staggered magnetizations (\ref{eq:LA}) are defined for tetrahedra of
type I.  Tetrahedra of type II become slaves: their magnetic state is
encoded in the staggered magnetizations of the four surrounding
tetrahedra of type I.  The vanishing of the total magnetization of
type-II tetrahedra yields the following constraint:
\begin{equation}
  \label{eq:phi3}
  \mathbf M^\mathrm{II} 
  = \phi_3\,\hat{\bf n}_3 - \partial_y \hat{\bf n}_1/4 = 0.
\end{equation}
From it we infer that spatial derivatives of $\hat{\bf n}_1$ are of
the same order as $\phi_2$ and $\phi_3$.  The N\'eel magnetizations
of a type-II tetrahedron are, to lowest orders,
\begin{eqnarray}
  \mathbf L^\mathrm{II}_1 = \phi_2\hat{\bf n}_2-\partial_z \hat{\bf n}_1/4,
  \quad
  \mathbf L^\mathrm{II}_2=\hat{\bf n}_1, 
  \quad
  \mathbf L^\mathrm{II}_3 = -\partial_x \hat{\bf n}_1/4.
  \label{eq:LB}
\end{eqnarray}

Upon adding contributions from tetrahedra of both types and using Eq. 
(\ref{eq:phi3}) we obtain the DM energy
\begin{equation}
  \label{eq:EDM}
  E_\mathrm{DM} = -DS^2 \, \hat{\bf n}_1 \cdot (
  \hat{\bf a} \times \partial_x \hat{\bf n}_1
  +\hat{\bf b} \times \partial_y \hat{\bf n}_1
  -\hat{\bf c} \times \partial_z \hat{\bf n}_1)/4.
\end{equation}
The terms linear in the spatial derivatives make a uniform state
unstable against the formation of a spiral \cite{Dzyaloshinskii}.  The
pitch of the spiral depends on the stiffness of the staggered
magnetization, which is ordinarily determined by the strength of
exchange.  However, the large degeneracy of the pyrochlore
antiferromagnet with nearest-neighbor exchange leads to a vanishing
stiffness: indeed, apart from the constraint (\ref{eq:phi3}), the
direction of $\hat{\mathbf n}_1$ can vary arbitrarily in space.  The
stiffness is therefore determined by the magnetoelastic coupling and
by weak exchange interactions beyond nearest neighbors.
We discuss the magnetoelastic coupling first.

A spiral magnetic state represents a deviation from the collinear
structure and thus increases the magnetoelastic energy.  On symmetry
grounds, the increase should be quadratic in the gradients of
$\hat{\mathbf n}_1$ and thus may yield a finite stiffness.  For
simplicity, we first consider only the odd distortions, effectively
setting $k_g = \infty$ and $K_g = 0$ in Eq.~(\ref{eq:Eme}), and
discuss the influence of the even phonon later.  The magnetoelastic
energy can then be expressed in terms of the odd doublet as $E_\mathrm{me} =
-K_u \mathbf u \cdot \delta \mathbf u/2$, where $\mathbf u =
4\,S^2 \,(0,1)$ is the value in the commensurate ground state
and $\delta \mathbf u$ is a small deviation.
As a result, we obtain the magnetoelastic energy density as a
function of $\phi_i$ and the gradients of $\hat{\mathbf n}_i$.
However, on account of the constraint (\ref{eq:phi3}), $\phi_3 =
\hat{\mathbf n}_3 \cdot \partial_y \hat{\mathbf n}_1/4$.  Likewise,
minimization of the energy with respect to $\phi_2$ yields
\begin{equation}
\phi_2 = \hat{\mathbf n}_2 \cdot \partial_z \hat{\mathbf n}_1/8.
\label{eq:phi2}
\end{equation}
The energy cost associated with the spiral is then
\begin{eqnarray}
  E_\mathrm{me} = K_u \, (S^4/4) \!\! && [ (\partial_x \hat{\bf n}_1)^2
    +(\partial_y \hat{\bf n}_1)^2 
    +2\,(\partial_z \hat{\bf n}_1)^2
    \nonumber\\
    &&-(\hat{\bf n}_2\cdot\partial_z \hat{\bf n}_1)^2 ].
  \label{eq:Eme3}
\end{eqnarray}

The total energy of a spiral state is the sum of the DM energy
(\ref{eq:EDM}) and the magnetoelastic energy (\ref{eq:Eme3}).  
Its minimization yields three simple
spiral states in which $\hat{\bf n}_1$ rotates about one of
the principal axes staying in the plane perpendicular to it, e.g.
$\hat{\bf n}_1 = (0, \cos{\theta(x)}, \sin{\theta(x)})$.  The energy 
density of all three states is the same,
\begin{equation}
\label{eq:Etot}
E = -DS^2 \theta' + K_u S^4 {\theta'}^2/4,
\end{equation}
where, in this case, $\theta'=\partial_x \theta$.  The pitch of the spiral is
\begin{equation}
 \theta' = 2\pi \delta =  2D/ S^2 K_u.
 \label{eq:delta}
\end{equation}

\begin{figure}
\includegraphics[width=0.99\columnwidth]{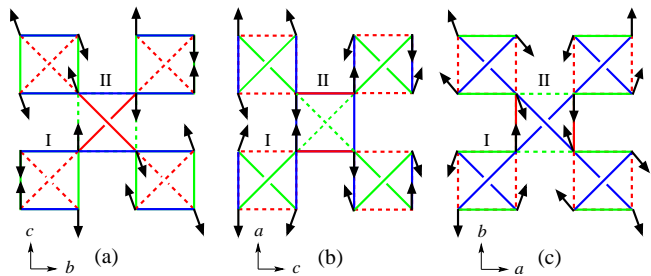}
\caption{(Color online)
Spiral magnetic orders with the spins rotating about (a) the $a$ axis, 
(b) the $b$ axis, and (c) the $c$ axis.}
\label{fig:spiral}
\end{figure}

A spiral state with spins rotating about the $a$ axis is shown in
Fig.~\ref{fig:spiral}(a).  On account of Eqs.~(\ref{eq:phi3}) and
(\ref{eq:phi2}), we have $\phi_2 = \phi_3 = 0$, so that 
type-I tetrahedra still have collinear spins.  The spins on type-II
tetrahedra are coplanar with the twist angle of $\theta'/4 = \pi
\delta/2$.  The spiral magnetic order produces Bragg scattering at 
$\mathbf q =  2\pi (-\delta,0,1)$.  

Fig.~\ref{fig:spiral}(b) shows a spiral state twisting
about the $b$ axis.  This time tetrahedra of type I have coplanar
spins with the twist angle $\phi_3 = \theta'/4 = \pi\delta/2$, while
tetrahedra of type II have collinear spins, as can be checked by using
Eq.~(\ref{eq:LB}).  This spiral is related to the previous
one by a lattice symmetry (the inversion and a $\pi/2$ rotation
in the $xy$ plane).  It produces a magnetic Bragg peak at 
$\mathbf q = 2\pi (0,\delta,1)$, as observed by Chung {\em
et al.} \cite{Chung05}.  The measured value $\delta\approx 0.09$ 
\cite{Chung05} is consistent with a DM interaction that is weak 
relative to the magnetoelastic coupling: $D/K_u S^2\approx0.28$.  

The third spiral solution is shown in Fig. \ref{fig:spiral}(c).  It
has the wavevector ${\bf q} = 2\pi(0,0,1+\delta)$; the spins are
rotating around the $c$-axis.  The twist angle is $\phi_2=\delta\pi/4$.

The degeneracy of the three spiral ground states is lifted when other
perturbations are taken into account.  A uniform $E_g$ distortion and 
the further-neighbor exchanges $J_2$ and $J_3$ add energy terms 
\begin{equation}
  K_g (S^4/4)[(\partial_x \hat{\bf n}_1)^2 
  + (\partial_y \hat{\bf n}_1)^2]
  + (4J_3-2J_2) S^2 (\partial_z \hat{\bf n}_1)^2.
\label{eq:perturb}
\end{equation}
Depending on these coupling constants, the system will prefer 
the spiral states shown either in Fig.~\ref{fig:spiral}(a) and (b) 
or in Fig.~\ref{fig:spiral}(c).

{\em Density-functional calculations.}  To test this theory,
we performed density-functional calculations within the LSDA+U
method \cite{Anisimov} using projector augmented-wave potentials 
as implemented in the {\it Vienna ab initio Simulation Package} 
\cite{VASP,PAW}.  Values for
the on-site Coulomb and exchange parameters, $U = 3$ eV and $J =
0.9$ eV, were choosen as previously described \cite{Fennie}. The
results are not particularly sensitive to reasonable variations of
$U$ ($\pm 1$ eV).  First we performed full structural relaxations in
the 14-atom cubic unit cell, space group $Fd\bar{3}m$. 
Chromium ions were initialized
with parallel spins in order to retain $O_h$ symmetry throughout the
structural relaxation. The relaxed lattice constant, $a$=8.54 \AA, was
found to be in excellent agreement with $a$=8.59 \AA\/ measured by
Chung {\em et al.} \cite{Chung05}. 

Estimates of exchange constants were obtained by comparing the total
energy for several simple spin configurations.  To prevent
contamination by magnetoelastic terms (\ref{eq:Eme}), the lattice
structure was frozen in the reference cubic state with $a=8.54$ \AA.
That procedure yielded $J_1 = 0.5$ meV, $J_2 \approx 0$ meV, and
$J_3 = 0.15$ meV.  The resulting Curie-Weiss temperature
$\Theta_\mathrm{CW}=-(1/3k_B) S(S+1) \sum_{i}z_iJ_i = -70$ K compares
well to the experimental values ranging from $-70$ to $-90$ K
\cite{Chung05,Ueda05}.

To quantify the magnetoelastic effects, we performed full structural
relaxations for three spin configurations: (i) a collinear state with
a pure $E_g$ distortion and magnetic wavevector $\mathbf q=0$ shown in
Fig.~5 of Ref.~\onlinecite{OT02b}; (ii) a coplanar state with a pure
$E_g$ distortion and $\mathbf q = 2\pi (0,0,1)$; and (iii) a collinear
state with a mixed $E_u + E_g$ distortion and $\mathbf q = 2\pi
(0,0,1)$ shown in Fig.~\ref{fig:redgreen}.  States (ii) and (iii) are
the commensurate limits ($\delta \to 0$) of the states displayed in
Figs. 3(c) and (d) of Ref.  \onlinecite{Chung05}; they have a doubled
unit cell (28 atoms).  The total energy
was lowest in the collinear state (iii), as posited above.  The $E_g$
component of the distortion is tetragonal.  Its value, $(c-a)/c = +5.1
\times 10^{-3}$, is remarkably close to the experimental one
\cite{Chung05}.  Reduction of the energy associated with the
structural relaxation depends on the spin state according to
Eq.~(\ref{eq:Eme}).  From the data obtained in the three reference
states we deduced the magnetoelastic constants $K_u = 0.15$ meV $ >
K_g = 0.13$ meV.  Quantitatively, the spin-lattice coupling is weaker 
than exchange, although not by much: $K_u S^2/J_1 = 0.67$.

The {\em ab initio} calculations back up the conclusions obtained
analytically.  The phonon doublet $E_u$ indeed turned out to be softer
than the even distortion $E_g$ confirming the selection of the
collinear state of Fig.~\ref{fig:redgreen}.  The competition between
the three candidate spiral states is decided by the ratio of the
coupling constants in Eq.~(\ref{eq:perturb}).  Because $J_3$ is quite
large, the last term is prohibitively expensive and the spiral twists
along either $x$ or $y$, as indeed observed \cite{Chung05}.

A large value of $J_3$ may cast
doubts on the applicability of Eq.~(\ref{eq:perturb}), which treats
that coupling as a small perturbation.  However, it turns out that our
conclusions in that regard remain valid for arbitrarily large values
of $J_3$.  Luckily for us, the spiral states depicted in
Fig.~\ref{fig:spiral}(a) and (b) minimize the third-neighbor exchange
exactly for any value of the pitch $\delta$ \cite{Chern07}.  The 
third spiral state increases the $J_3$ term and is thus suppressed.

To appreciate the unusual microscopic origin of the magnetic spiral,
it is helpful to compare it to the standard scenario exemplified by
the ferroelectric BiFeO$_3$ \cite{Kadomtseva04}.  In the latter case,
the pitch of the magnetic spiral is proportional to an order parameter
measuring the violation of parity, such as the electric dipolar moment.  
In contrast, in CdCr$_2$O$_4$ the 
strongest violation of parity comes not so much from the lattice distortion 
as from the magnetically ordered state itself: frustrated bonds form spirals of
the same handedness (Fig.~\ref{fig:redgreen}).  The corresponding
order parameter---the odd doublet of bond variables $\mathbf u$
\cite{OT02b}---is a dimensionless quantity of order 1, which is why
the presence of an order parameter is not immediately evident in
Eq.~(\ref{eq:delta}).  Without parity violation, the
DM interaction alone would not generate a magnetic helix.  

We thank C. Broholm, D. Clarke, H. D. Drew, C. L. Henley, S.-H. Lee,
R. Moessner, and A. B. Sushkov for stimulating discussions.  This work
was supported in part by the NSF Grant No.  DMR-0348679.

\end{document}